# Adaptive Hybrid Deflection and Retransmission Routing for Optical Burst-Switched Networks

Martin Lévesque, Halima Elbiaze
Department of Computer Science
Université du Québec à Montréal
Montréal (QC), Canada
Email: elbiaze.halima@uqam.ca

Wael Hosny Fouad Aly
Department of Computer Engineering
Arab Academy for Science & Technology
Alexandria, Egypt
Email: drwaelhosny@aast.edu

*Abstract*—Burst contention is a well known challenging problem in *Optical Burst Switching* (OBS) networks. Deflection routing is used to resolve contention. Burst retransmission is used to reduce the *Burst Loss Ratio* (BLR) by retransmitting dropped bursts. Previous works show that combining deflection and retransmission outperforms both pure deflection and pure retransmission approaches. This paper proposes a new *Adaptive Hybrid Deflection and Retransmission* (AHDR) approach that dynamically combines deflection and retransmission approaches based on network conditions such as BLR and link utilization. *Network Simulator 2* (ns-2) is used to simulate the proposed approach on different network topologies. Simulation results show that the proposed approach outperforms static approaches in terms of BLR and goodput.

## I. INTRODUCTION

*Optical Burst Switching* (OBS) [1] is a promising technology to handling bursty and dynamic Internet Protocol traffic in optical networks effectively.

In OBS networks, user data (IP for example) is assembled as a huge segment called a *data burst* which is sent using *one-way resource reservation*. The burst is preceded in time by a control packet, called *Burst Header Packet* (BHP), which is sent on a separate control wavelength and requests resource allocation at each switch. When the control packet arrives at a switch, the capacity is reserved in the cross-connect for the burst. If the needed capacity can be reserved at a given time, the burst can then pass through the cross-connect without the need of buffering or processing.

Since data bursts and control packets are sent out without waiting for an acknowledgment, the burst could be dropped due to resource contention or to insufficient offset time if the burst catches up the control packet. Thus, it is clear that burst contention resolution approaches play an essential role to reduce the *Burst Loss Ratio* (BLR) in OBS networks [2].

Burst contention can be resolved using several approaches, such as *wavelength conversion*, *buffering* based on *fiber delay line* (FDL) or *deflection routing*. Another approach, called *burst segmentation*, resolves contention by dividing the contended burst into smaller parts called *segments*, so that a segment is dropped rather than the entire burst.

Deflection routing is the most attractive solution to resolve the contention in OBS networks, because it does not need added cost in terms of physical components and uses the available spectral domain. However, as the load increases, deflection routing could lead to performance degradation and network instability. Since deflection can not eradicate the burst loss, retransmission at the OBS layer has been suggested by Torra et al. [3].

A static combination of deflection and retransmission has been proposed by Son-Hong Ngo et. al. [4]. They have proposed a *Hybrid Deflection and Retransmission* (HDR) algorithm [4] which combines deflection routing and retransmission. Simulation results have shown that HDR gives bad overall performance because it systematically try deflection first. To overcome this shortcoming, the authors have developed another mechanism called *Limited Hybrid Deflection and Retransmission* (LHDR) that limits the deflection.

This paper introduces a novel algorithm to combine deflection routing and retransmission called *Adaptive Hybrid Deflection and Retransmission* (AHDR). A success probability threshold function is used to dynamically make the decision of using either the deflection or the retransmission based on local knowledge about network conditions. In order to make this local knowledge feasible, AHDR algorithm exploits sending and receiving of *Positive Acknowledgement* (ACK) and *Negative Acknowledgement* (NACK) messages to advertize useful statistics about the network conditions stored by all nodes.

This paper is organized as follows. Section II describes the proposed Adaptive Hybrid Deflection and Retransmission (AHDR) algorithm. Section III presents simulation results. Finally, Section IV contains the conclusion and future possible works.

## II. ADAPTIVE HYBRID DEFLECTION AND RETRANSMISSION

In this section, we describe the proposed algorithm (AHDR). AHDR optimizes the decision of doing either a deflection or a retransmission. It also enhances the selection of an alternate route.

### A. Transferring statistics between nodes

Once the control packet reaches the destination, an ACK is sent to the source. If the control packet is dropped, then the proposed algorithm uses a NACK to notify the source for burst retransmission. AHDR does not only use the ACK and the NACK for notification but it uses them also to transmit some statistics about links states (Fig. 1).

Indeed, the BLR and the utilization are measured on each link and this information is integrated into the ACK packets. In the case of NACK, statistics are collected by using the link between the current node and the next node of the route. In the case of ACK, the BLR and the utilization between the destination node and the last node before the destination are used. When a node receives an ACK or a NACK control packet, this node collects and analyzes statistics. Thus, statistics of the whole network are eventually updated.

### B. Success probability calculation

First, to limit the length of deflection routes, we introduce a parameter (noted $\xi$) which expresses the deflection route length threshold. Let $Defl$ denotes a possible deflection route, $Primary$ the primary route and $|Route|$ the number of hops of the route $Route$. If $|Defl| <= |Primary| * \xi$, then $Defl$ is added as a possible deflection alternative.

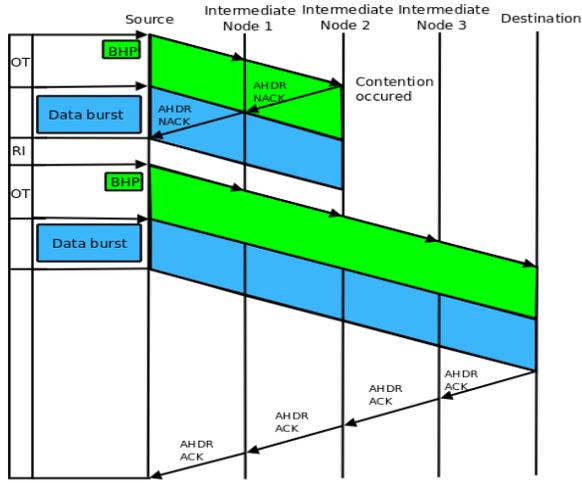

Fig. 1. Signaling scheme used by AHDR with a retransmitted burst

Second, AHDR incorporates BLR and utilization weights to measure the *dropping probability* (DP) between two nodes as follows:

$$DP(n_1, n_2) = W_{BLR} * BLR_{(n_1,n_2)} + W_U * U_{(n_1,n_2)} \quad (1)$$

where $U$ is the utilization, $n_1$ and $n_2$ are two adjacent nodes, $DP$ returns the dropping probability between $n_1$ and $n_2$, $W_{BLR}$ is a weight applied to the BLR and $W_U$ is another weight applied to the utilization. We note that $W_{BLR} + W_U = 1$.

The success probability ($SP(R)$) of a route $R$ is defined as follows:

$$SP(R) = \prod_{i=1}^{|R|-1} (1 - DP(n_i, n_{i+1})) \quad (2)$$

The success probability of the link between $n_i$ and $n_{i+1}$ is given by $1 - DP(n_i, n_{i+1})$. Eq. 2 multiplies all success probability links to get a global success probability for the entire route.

In AHDR, a success probability threshold is defined to make the decision of either deflecting or retransmitting a given burst in contention. In order to take into account the network conditions, we introduce a dynamic threshold function $SP_{th}(BLR)$:

$$SP_{th}(BLR) = \omega * BLR + \varphi \quad (3)$$

where $\omega$ is the slope of the function and $\varphi$ the intercept. The algorithm computes a regression line [5] with minimum Burst Loss Ratios associated with the best success probability threshold to use.

If the success probability (Eq. 2) of a given deflection alternative is greater or equal than the associated threshold (Eq. 3), then it means that this alternative should currently be tried.

Obviously, those formulas are pre-calculated and a typical routing table is periodically updated so that the forwarding process is not penalysed.

### C. AHDR algorithm

Fig. 2 shows AHDR algorithm.

- When a control packet is received, the current node is compared to the destination node. If the BHP arrives at the destination, then an ACK is sent to the source.

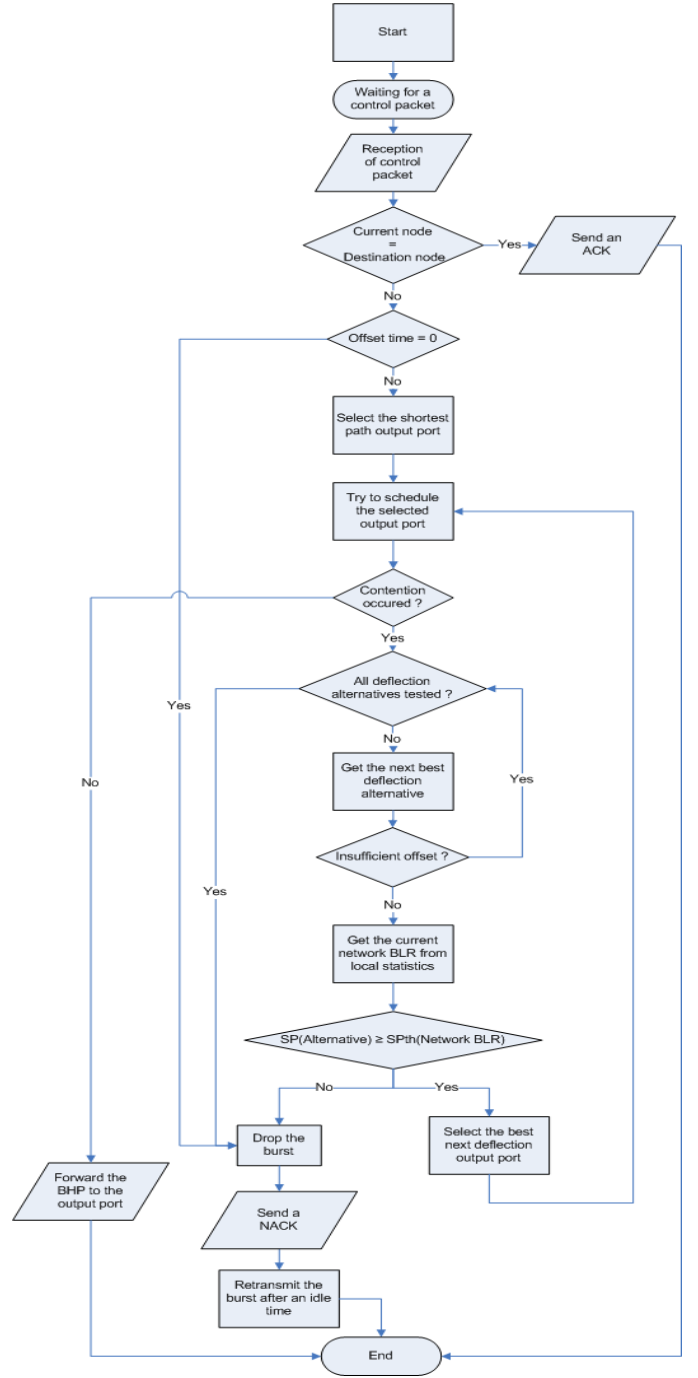

Fig. 2. Flowchart of forwarding process

- Then, the offset time is checked in order to verify if it is still sufficient. If it is not sufficient, a NACK is sent to the source and the burst is retransmitted after an idle time.
- The shortest path output port is selected. In case of resource contention, it is solved by either deflecting or by retransmitting the burst.
- AHDR successively extracts the best deflection alternatives in order to minimize the BLR and the number of retransmissions. The best output port is found by extracting the next hop in the route $R$ where $SP(R)$ (Eq. 2) is the greatest.

- The success probability of the deflection route is then compared to a dynamic success probability threshold.
- If the success probability of the current alternative is smaller than the threshold, then a NACK is sent to the source and the burst is retransmitted after an idle time. Otherwise, the current output port alternative is scheduled.

### D. Adaptive offset time

In OBS networks, data burst follows the control packet after a predetermined offset time calculated at the ingress node. The offset time has to be large enough so that bursts arrive at each switch after the control packet. The minimum offset time $t_{offset}$ must considerates the BHP processing time $t_p$ at each hop, the node switching and the configuration time $t_{conf}$. However, the minimum offset time is expressed by:

$$t_{offset} = t_{conf} + N_{hops} * t_p \quad (4)$$

where $N_{hops}$ is the number of hops. Eq. 4 expresses the fact that the main key to find the best offset time is to predict the number of hops because $t_{conf}$ and $t_p$ are fixed. However, if deflection occurs, a longer route could be used, which may increases $N_{hops}$.

Let $DeflPermitted$ denotes a boolean variable to predict if the BHP will need a deflection or not. $DeflPermitted$ is true when $SP(defl) \geq SP_{th}(BLR)$. The number of hops ($N_{hops}$) to be used in the offset time equation (Eq. 4) is given by:

$$N_{Hops} = \begin{cases} |Max\ defl| & \text{if } DeflPermitted \text{ is true} \\ |Shortest\ path| & \text{otherwise} \end{cases} \quad (5)$$

where $|Max\ defl|$ means the path length of the longest deflection alternative from the ingress node. Eq. 4 is then used to calculate the offset time with an adapted number of hops. If $DeflPermitted$ is true, then the longest deflection route is used for the number of hops and otherwise the shortest path is used.

## III. SIMULATION RESULTS

This section shows a comparison between AHDR and LHDR. Simulations are performed with *NSF Network* (NSFNET) topology by using Network Simulator 2 (ns-2) with an extra module for OBS. To evaluate the performance of AHDR, two different scenarios were investigated:

- General scenario: NSFNET with a total of $\tau$ traffic generators distributed from all nodes to all nodes.
- Bottleneck scenario: NSFNET with a total of $\tau$ traffic generators distributed only on seven random selected nodes. Traffic generators then send bursts to any selected node.

The following simulation configuration is used:
- Each wavelength has 1 Gbit/s of bandwidth capacity.
- Each link has 2 control channels and 4 data channels.
- The mean burst size equals 400 KB.
- Packet generation follows an exponential distribution for the burstification rate and the burst size.
- Bursts are indefinitely lost after a certain number of retransmissions $N_{ret}$ (truncated retransmission). $N_{ret}$ is fixed to 1 in order to not increase significantly the end-to-end delay. Finding the best $N_{ret}$ is out of the scope of this paper.
- We define the traffic load to be the ratio of the total input source nodes throughput over the capacity to be used [6].

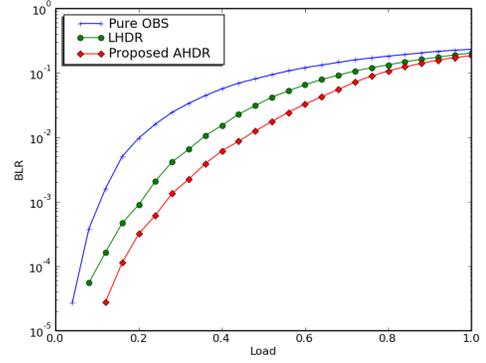

Fig. 3. Burst Loss Ratio (General scenario)

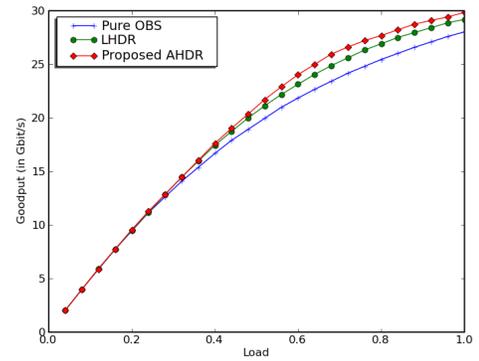

Fig. 4. Goodput (Gbit/s) (General scenario)

### A. Comparison of LHDR and AHDR

In this section, we present the obtained simulation results that compare LHDR and AHDR performance as well as a scenario given by OBS network without deflection or retransmission mechanisms (called Pure OBS). Let us recall that LHDR [4] uses the shortest path and a simple threshold function to limit the deflection whereas AHDR uses several parameters, a threshold function for the decision between a deflection or a retransmission and exploits the ACK and the NACK to transfer information about the network conditions in terms of BLR and link utilization.

In the general scenario, AHDR gives improvements compared to LHDR. Improvements are explained by the fact that the decision between doing a deflection or a retransmission is more effective (Fig. 3). Because of its adaptive characteristic, AHDR makes more deflections when the load is low and limits the deflection as the load increases. Plus, as the traffic increases, AHDR changes the decision adaptively by issuing more retransmissions. It also continues to do deflections with low loaded links. That is why it does not converge to the LHDR curv. Fig. 4 illustrates the goodput (expressed in Gbit/s) with the general scenario. It is clearly shown that AHDR achieves a higher goodput because decisions to resolve contention are made more efficiently.

In the bottleneck scenario, the obtained results in terms of BLR and goodput cleary shows that AHDR outperforms LHDR (Fig. 5 and 6). AHDR performs a lot of effective deflections, compared to LHDR which always uses the shortest path where the decision is made by static metrics like the number of hops. If we compare Pure OBS with LHDR with high network load (load $\geq 0.8$), we can see that LHDR becomes nearly ineffective and gives almost no improvement in terms of BLR and goodput. However, optimizing decisions (between doing a deflection of a retransmission) gives significant improvement even at high network load.

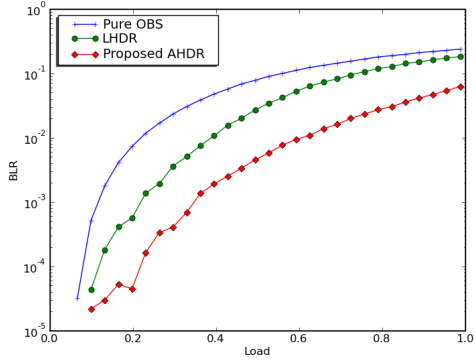

Fig. 5. Burst Loss Ratio (Bottleneck scenario)

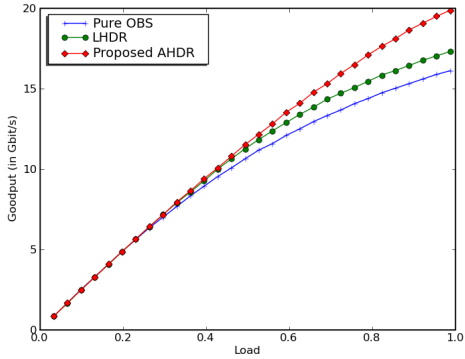

Fig. 6. Goodput (Gbit/s) (Bottleneck scenario)

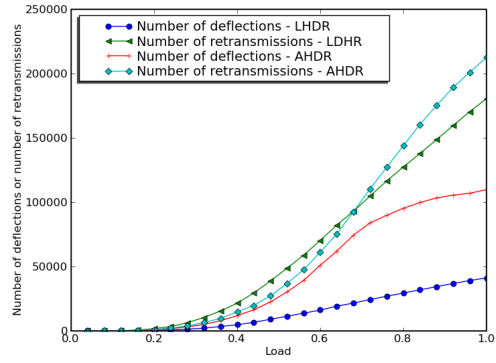

Fig. 7. Number of deflections versus number of retransmissions (General scenario)

To summarize this first set of results, we can conclude that when the traffic is uniformly distributed in the network, AHDR gives significative improvement compared to static approaches like LHDR. But when the traffic is not uniformly distributed in the network, AHDR outperforms static approaches because it selects low loaded links and makes dynamic decisions among several contention resolution strategies (deflection and retransmission in this paper).

### B. Number of deflections and number of retransmissions

Captures have been done in order to observe the number of deflections and the number of retransmissions while the load increases (Fig. 7). AHDR does more deflections and less retransmissions when the load is low and as the load increases it slowly reduces deflections. For the number of retransmissions, we can see that even if AHDR does more deflections, the number of retransmissions does not increase significantly comparing to LHDR which basically means that most of the deflections are effective.

## IV. CONCLUSION AND FUTURE WORKS

This paper presents a novel algorithm called *Adaptive Hybrid Deflection and Retransmission* (AHDR) that combines deflection and retransmission routing. The decision is taken based on parameters such as BLR and link utilization. For an effective decision to do a deflection or a retransmission, AHDR uses a success probability threshold calculated dynamically with collected statistics. Results show that AHDR is much more stable than static algorithms which don't consider the traffic conditions. The strength of AHDR is that it adapts decisions dynamically in order to resolve contentions. At low load, AHDR performs more deflections. At higher load, AHDR reduces the number of deflections and increases the number of retransmissions to decrease the BLR. Selecting the route to do a deflection to is more effective than using always the shortest path since AHDR selects the least congested route and also discovers links having low utilization.

The future work of this research is to combine several contention resolution strategies in a dynamic way because we believe that the feasibility of OBS requires effective and adaptive algorithms to overcome the burst loss issue. We are presently working on a new approach which deploys a probabilistic graphical model used in artificial intelligent in order to make efficient and dynamic decisions among several contention resolution strategies.

### ACKNOWLEDGMENT

The authors would like to thank Son-Hong Ngo for answering several of our questions about his paper [4].